\documentclass[apj]{emulateapj}

\usepackage[colorlinks=true,allcolors=blue]{hyperref} 
\def\R200{$R_{200}$}
\def\Msun{M$_{\odot}$}
\def\oii{[O\,{\scriptsize II}]}

\shorttitle{Molecular gas reservoirs in $z=1.46$ cluster galaxies}
\shortauthors{Hayashi M. et al.}

\begin{document}

\title{Molecular gas reservoirs in cluster galaxies at z=1.46}

\author{Masao Hayashi\altaffilmark{1},
  Ken-ichi Tadaki\altaffilmark{1},
  Tadayuki Kodama\altaffilmark{2},
  Kotaro Kohno\altaffilmark{3,4},
  Yuki Yamaguchi\altaffilmark{3},  
  Bunyo Hatsukade\altaffilmark{3},  
  Yusei Koyama\altaffilmark{5,6},
  Rhythm Shimakawa\altaffilmark{1,7},
  Yoichi Tamura\altaffilmark{8},
  Tomoko L. Suzuki\altaffilmark{1}
}
\affil{$^{1}$National Astronomical Observatory of Japan, Osawa, Mitaka, Tokyo 181-8588, Japan; masao.hayashi@nao.ac.jp} 
\affil{$^{2}$Astronomical Institute, Tohoku University, Aramaki, Aoba-ku, Sendai 980-8578, Japan}
\affil{$^{3}$Institute of Astronomy, Graduate School of Science, The University of Tokyo, 2-21-1 Osawa, Mitaka, Tokyo 181-0015, Japan}
\affil{$^{4}$Research Center for the Early Universe, Graduate School of Science, The University of Tokyo, 7-3-1 Hongo, Bunkyo-ku, Tokyo 113-0033, Japan}
\affil{$^{5}$Subaru Telescope, National Astronomical Observatory of Japan, 650 North A'ohoku Place, Hilo, HI 96720, USA}
\affil{$^{6}$Department of Astronomical Science, SOKENDAI (The Graduate University for Advanced Studies), Mitaka, Tokyo 181-8588}
\affil{$^{7}$UCO/Lick Observatory, University of California, 1156 High Street, Santa Cruz, CA 95064, USA}
\affil{$^{8}$Department of Physics, Nagoya University, Furo-cho, Chikusa-ku, Nagoya 464-8601, Japan}  

\begin{abstract}
We present molecular gas reservoirs of eighteen galaxies associated
with the XMMXCS J2215.9-1738 cluster at $z=1.46$. From Band 7 and Band
3 data of the Atacama Large Millimeter/submillimeter Array (ALMA), we
detect dust continuum emission at 870 \micron\ and CO $J=$ 2--1
emission line from 8 and 17 member galaxies respectively within a
cluster-centric radius of \R200.  
The molecular gas masses derived from the CO and/or dust continuum
luminosities show that the fraction of molecular gas mass and the
depletion time scale for the cluster galaxies are larger than expected
from the scaling relations of molecular gas on stellar mass and offset
from the main sequence of star-forming galaxies in general fields. 
The galaxies closer to the cluster center in terms of both
projected position and accretion phase seem to show a larger deviation
from the scaling relations. We speculate that the environment of
galaxy cluster helps feed the gas through inflow to the member
galaxies and also reduce the efficiency of star formation.        
The stacked Band 3 spectrum of 12 quiescent galaxies with 
$\rm M_{stellar}\sim10^{11}$ \Msun\ within 0.5\R200\ shows no
detection of CO emission line, giving the upper limit of molecular gas
mass and molecular gas fraction to be $\la10^{10}$ \Msun\ and $\la10$\%,
respectively. Therefore, the massive galaxies in the cluster core
quench the star formation activity while consuming most of the gas
reservoirs. 
\end{abstract}



\section{Introduction}

Quiescent galaxies dominate galaxy clusters in the local Universe
\citep[e.g.,][]{Dressler1997,Peng2010,Scoville2013}, which implies
that the environment of galaxy clusters has an impact on the
transition of star-forming galaxies to quiescent galaxies.
Based on the Kennicutt--Schmidt relation
\citep{Schmidt1959,Kennicutt1998ApJ,Kennicutt2012} demonstrating
empirically that the gas content of galaxies is one of the most
essential quantities that govern star-formation activities in
galaxies, better understanding the evolution of galaxies in galaxy
clusters in terms of both star-formation activity and gas content
leads to identifying environmental processes responsible for quenching
of star formation in galaxies.       

Most of star-forming galaxies in the local Universe follow a tight
positive correlation between star formation rate (SFR) and stellar
mass which is called a main sequence of star-forming galaxies
\citep[e.g.,][]{Daddi2007,Elbaz2007,Noeske2007,Renzini2015}.  
Gas fraction and star formation efficiency of galaxies can be
responsible for deviation from the main sequence in the plane of
SFR-M$_{\rm stellar}$, in the sense that starburst (passive) galaxies
tend to have larger (smaller) gas fraction and/or higher (lower)
efficiency of star formation \citep{Saintonge2012,Saintonge2016,Saintonge2017,Sargent2014}.   
On the other hand, as long as we focus on star-forming galaxies, an
environment where galaxies reside do not have a strong impact on the
star-forming main sequence and the relationship between gas reservoirs
and star-formation activity \citep{Peng2010,KoyamaS2017}. These
observational studies may suggest that star formation activity in most
of galaxies is governed not by external process such as galaxy
interaction but by internal factors such as gas reservoir.  
However, since galaxies in the local clusters have already evolved, it
is essential to investigate evolving cluster galaxies in the early
Universe to reveal how the present-day quiescent galaxies quench the
star formation within the galaxy clusters.  

Observations in the high-$z$ Universe have also been conducted
actively and it is found that a main sequence of star-forming galaxies
exists at each redshift up to $z\sim3$ or higher \citep[e.g.,][]{Speagle2014,Whitaker2014,Schreiber2015}.  
Now that it becomes possible to compile about three orders of
magnitude measurements of molecular gas from individual galaxies and
stacks at $z=$ 0--4, scaling relations of molecular gas fraction and
depletion time scale on offset from the main sequence, i.e., stellar
mass, SFR, and redshift, are constructed \citep{Genzel2015,Tacconi2018,Scoville2017}. 
The gas fraction of galaxies tend to be larger at higher redshifts 
\citep[e.g.,][]{Tacconi2010,Tacconi2013,Geach2011,Saintonge2013,Scoville2017}, 
as if it follows the redshift evolution of cosmic SFR density
\citep{Madau2014}, suggesting that the SFR of a galaxy with a given
mass becomes larger in proportion to the gas fraction as the redshift
increases. In spite of the remarkable recent progress, most of the
observations of molecular gas at high redshifts have been limited to
the galaxies in general fields 
\citep[e.g.,][]{Magnelli2012b,Carilli2013,Walter2014,Genzel2015,Silverman2015,Decarli2016,Decarli2016b,Seko2016,Tacconi2018,Scoville2017}.   
An increasing number of studies have surveyed molecular gas in galaxy
(proto-)clusters at high redshifts of $z\approx$ 1--3, however, the
measurements of the gas content are at most for a few member galaxies
in each cluster \citep{Wagg2012,Aravena2012,Casasola2013,Ivison2013,Tadaki2014b,Chapman2015,Wang2016,Dannerbauer2017,Noble2017,Rudnick2017,Stach2017,Lee.M2017,Webb2017}.      

XMMXCS J2215.9-1738 galaxy cluster at $z=1.457$ 
\citep[22$^{\rm h}$15$^{\rm m}$58$^{\rm s}$.5, -17$^\circ$38$'$02.5$''$;][]{Stanford2006} 
is one of the best targets to probe the early phase of
environmental effects on molecular gas properties in cluster
galaxies. This is because in addition to previous studies indicating
that massive galaxies in the cluster core are still in their formation
phase \citep{Hayashi2010,Hayashi2014,Hilton2010,Ma2015},
CO $J=$ 2--1 ($\nu_{\rm rest}$ = 230.538 GHz, hereafter CO(2--1))
emission lines are found with ALMA from 17 galaxies associated with
the galaxy cluster \citep{Hayashi2017}. Accretion phases of the gas-rich 
member galaxies are discussed based on the phase space of relative
velocity versus cluster-centric distance. The galaxies with CO(2--1)
detected disappear from the very center of the cluster, suggesting
that the gas-rich galaxies have entered the cluster more recently than
the gas-poor galaxies located in the virialized region of this cluster. 
\citet{Hayashi2017}, for the first time, succeed in detecting
CO(2--1) emission lines from as many as 17 member galaxies in the
cluster at $z = 1.46$. Next step to better understanding of the
evolution of cluster galaxies is to investigate their gas reservoirs
and efficiency of star formation. 

\citet{Stach2017} have independently detected fourteen 1.25mm dust
continuum sources from their own ALMA data in the central region of
the XMMXCS J2215.9-1738 cluster. Among them, the eleven sources are
confirmed to be cluster members and the six sources have both CO(2--1)
and CO(5--4) emission lines detected. The detections of dust continuum
and/or CO line are consistent with those reported by
\citet{Hayashi2017}. The ratio of CO luminosities from the different
transitions in the cluster is similar to those for field galaxies at
similar redshifts. Gas masses of $\sim$(1--2.5)$\times10^{10}$
\Msun\ and a relatively short gas consumption timescale of $\sim$200
Myr are estimated for the galaxies under assumption of a conversion
factor of $\alpha_{CO}=1$. They argue that based on the line widths
and luminosities of the two CO transitions, the CO(2--1) gas tends to
be stripped from the galaxies rather than the CO(5--4) gas, which
implies an environmental process acts on the cluster galaxies.

In this paper, we present full discussions from our ALMA observations
in Band 3 and Band 7 in the XMMXCS J2215.9-1738 galaxy cluster through
two programs of 2015.1.00779.S and 2012.1.00623.S. The data in Band 3
and Band 7 allow us to detect CO(2--1) emission and dust continuum
emission at 870 \micron\ from cluster member galaxies,
respectively. We use the ALMA data to investigate molecular gas
reservoirs in the member galaxies and then discuss the evolution of
their star formation activities in terms of star formation efficiency
and gas consumption.
The outline of this paper is as follows. 
In Section \ref{sec:data}, the ALMA data as well as ancillary data
covering optical to mid-infrared (MIR) are described. The source
detection in the ALMA Band 7 data is performed and the photometric
catalog with multi-band photometry is created. 
In Section \ref{sec:results}, we derive molecular gas mass from the
ALMA data, and discuss fraction of molecular gas mass and depletion
time scale for the cluster member galaxies in the central region. In
Section \ref{sec:discussion}, we compare our results with the scaling
relation for field galaxies and results of other clusters at
$z\sim1.6$ from literature. We also discuss the molecular gas mass in
quiescent galaxies in the very center by stacking the Band 3 data. 
Conclusions are shown in Section \ref{sec:conclusion}.  
Throughout the paper, the cosmological parameters of $H_0=70$ km
s$^{-1}$ Mpc$^{-1}$, $\Omega_m=0.3$, and $\Omega_\Lambda=0.7$, along
with \citet{Chabrier2003} initial mass function (IMF), are adopted.  
The velocity dispersion of the cluster member galaxies is $\sigma=720$
km s$^{-1}$ and the radius of the galaxy cluster is \R200\ = 0.8 Mpc
\citep{Hilton2010}.

\begin{figure*}
  \begin{center}
    \includegraphics[width=\textwidth, clip, trim=190 0 350 150]{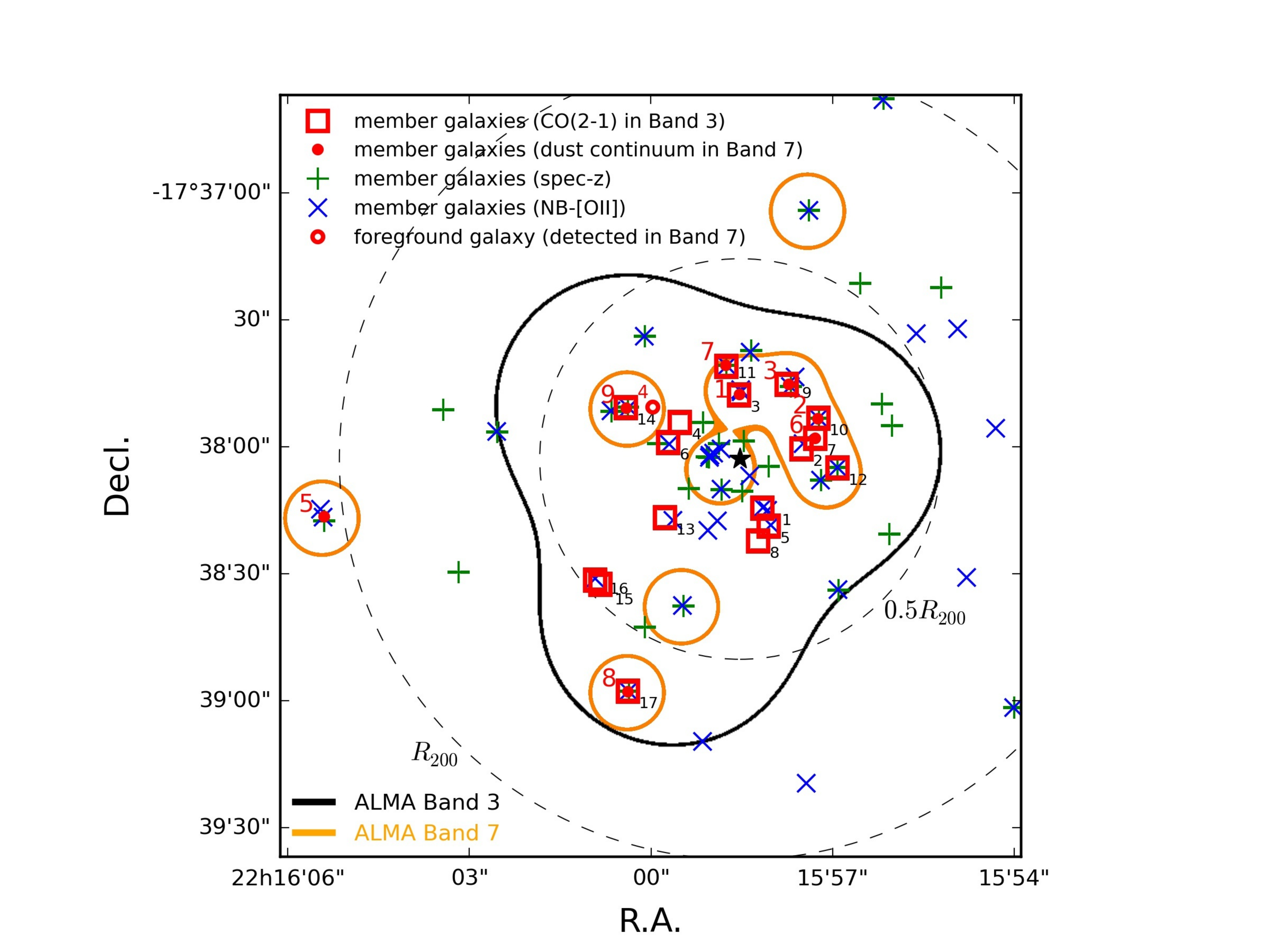}        
    \caption{
      Spatial distribution of galaxies detected in the ALMA
      data. Filled red circles show the cluster member galaxies with
      dust continuum detected in Band 7. Open red circle also shows a
      Band 7 source, but it is a foreground galaxy at $z=1.30$
      \citep{Stach2017}. Open squares show the member galaxies with
      detection of CO(2-1) emission line. The numbers next to the
      symbols are IDs shown in Tables~\ref{tbl:B7sources} and
      \ref{tbl:ALMAsources}. The black (orange) curves show the area
      where Band 3 (7) data are available. The green and blue crosses
      show the member galaxies confirmed by spectroscopy and
      \oii\ emitters associated with the cluster, respectively
      \citep{Hilton2010,Beifiori2017,Hayashi2014}. 
      A star symbol shows a cluster center determined with extended
      X-ray emission \citep{Stanford2006}. The dashed circles show the
      cluster-centric radius of $0.5R_{200}$ and $R_{200}$
      \citep{Hilton2010}.  
      \label{fig:map}
    }
  \end{center}
\end{figure*}

\begin{figure*}
  \begin{center}
    \includegraphics[width=\textwidth, clip, trim=30 20 30 10]{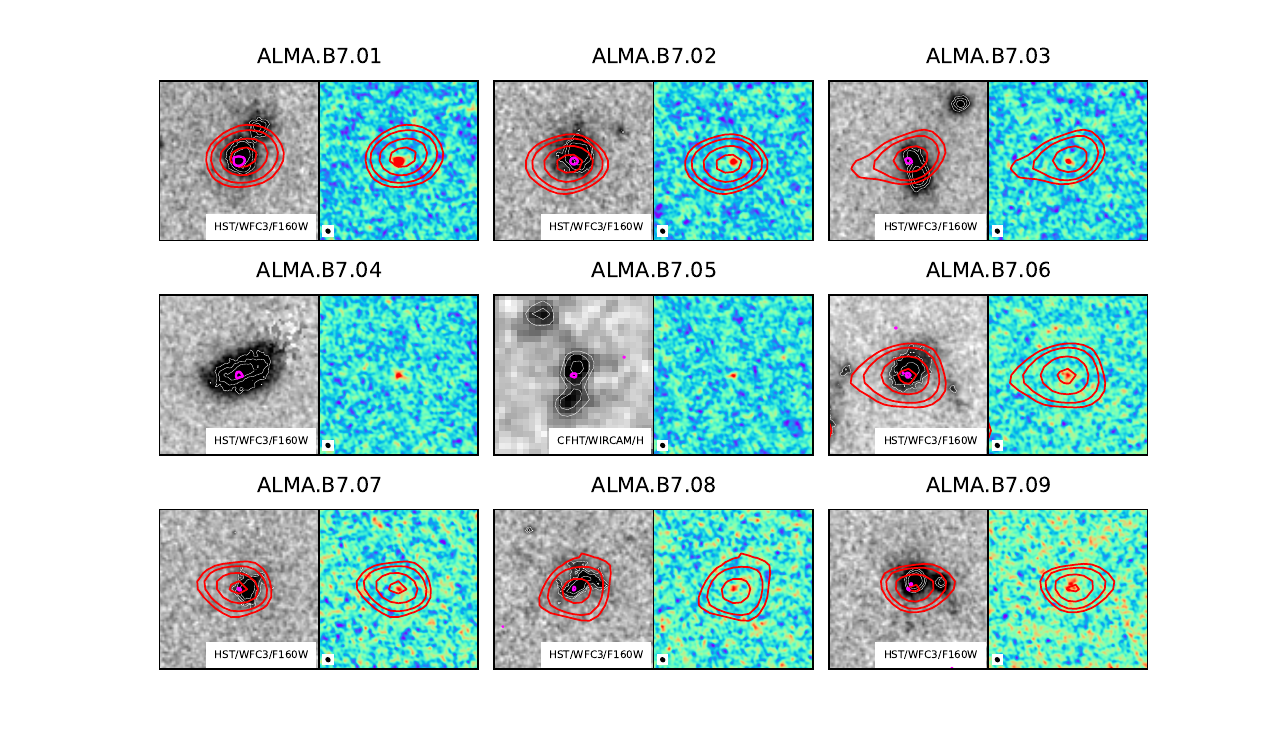}        
    \caption{
      Postage-stamps of the galaxies detected in ALMA Band 7, where
      5 arcsec on a side. The right panel shows the intensity map in
      Band 7, and the synthesized beam size is shown in black in the
      lower left corner. The left panel shows the $H$-band image
      (HST/WFC3 F160W data if available, otherwise CFHT/WIRCam
      $H$-band) along with the white contours of 2$\sigma$, 3$\sigma$,
      and 5$\sigma$ levels. It is also overlaid with the magenta
      contours showing the Band 7 intensity map of 4.0$\sigma$,
      4.5$\sigma$, and 5.0$\sigma$ levels. The red contours in both
      panels show the intensity of CO(2--1) emission lines
      \citep{Hayashi2017}. Note that the ALMA.B7.04 is not a cluster
      member galaxy but a foreground galaxy
      (Table~\ref{tbl:B7sources}). 
      \label{fig:b7images}
    }
  \end{center}
\end{figure*}

\section{Data}
\label{sec:data}

\subsection{ALMA Band 3}
\label{sec:data.band3}

\citet{Hayashi2017} already report the initial results from the Band
3 data. Since the details of the Band 3 data and source detection in
the data are described in the paper, we briefly mention them in this 
section. 

The Band 3 data are available in 2.33 arcmin$^2$ observed at three
pointings, where an area within 0.5 $R_{200}$ from the cluster center
is almost covered. The spectral coverage is 93.03 -- 94.86 GHz with a
spectral resolution of 13.906 MHz ($\sim$ 12.5 km
s$^{-1}$). Integration time is 1.04 hours per each pointing. Typical
noise level of the mosaicked 3D cubes is 0.12 mJy beam$^{-1}$ at
velocity resolutions of 400 km s$^{-1}$. The synthesized beam size is
1.79$''$ $\times$ 1.41$''$.   

We use {\tt Clumpfind} \citep{Williams1994} to search for emission
lines on the data cube. The emission-line search is performed in the
cubes with different velocity resolutions of 50, 100, 200, 400, and
600 km s$^{-1}$. We have detected 21 candidates at signal-to-noise
ratio (SNR) of $>5.0$ in at least one velocity resolution, after
excluding overlaps. We cross-match the detections in the ALMA data
with the optical and near-infrared (NIR) data described in
\S~\ref{sec:ArchiveData} to remove the possible false
detections. Consequently, we conclude that 17 emission lines with
counterparts in the optical--NIR data are secure detections. The
spectroscopic redshifts (if any), photometric information such as
colors are fully consistent with the counterparts being the member
galaxies. 
Note that the remaining four candidates are not detected in the 1.25mm
dust continuum data in ALMA Band 6 shown by \citet{Stach2017} and our
ALMA Band 7 data described below are not available to them.
The spectra, the intensity maps, and the properties of the
individual 17 emission lines are shown in \citet{Hayashi2017}. 
For cluster member galaxies without individual CO detection
(\S~\ref{sec:members}), we estimate an upper limit of CO luminosity
from $5\sigma$ noise level in the cube with 400 km $s^{-1}$ velocity
resolution.

\subsection{ALMA Band 7}
\label{sec:data.band7}

The observations in Band 7 were conducted in 2015 July. Four spectral
windows are set at central frequencies of 338, 340, 350 and 352 GHz
with each bandwidth of 1.875 GHz, respectively. The data are taken at
ten pointings to target 13 \oii\ emission-line galaxies
\citep{Hayashi2010} that have dust-corrected SFR$_{\rm \oii}$ of $>93$
\Msun\ yr$^{-1}$, which results in the patchy data coverage of 0.61
arcmin$^2$ (Figure~\ref{fig:map}). Integration time is 7.06 minutes
per each pointing. The synthesized beam size is 0.181$''$ $\times$
0.157$''$ with a position angle of 44.7 degrees. 
The spatial resolution is comparable to optical -- NIR data observed
with Hubble Space Telescope (HST). 

Calibration of the raw data is conducted using the Common Astronomy
Software Applications \citep[CASA version 4.3.1;][]{McMullin2007} with
a standard ALMA pipeline. The Briggs weighting with the robust
parameter of 2.0 (i.e., natural weighting) and a CLEAN threshold of
0.35 mJy ($\sim5\sigma$) are adopted to make CLEANed images. Among ten
pointings, the data at five pointings nearest from the cluster center
are mosaiced to make a single image (Figure \ref{fig:map}). Typical
noise levels of these images are 0.061 -- 0.068 mJy beam$^{-1}$, which
are measured by fitting a Gaussian to the histogram of pixel counts
while ignoring the bright end of the histogram that some bright sources
can contribute to.  

We search for sources with a pixel count larger than $4.6\sigma$ in
each image. We have detected 9 sources at 870\micron\ from the Band 7
data, all of which have a counterpart in optical and NIR data (Figure
\ref{fig:b7images}).  
Similar to the 1.25mm dust continuum sources reported by
\citet{Stach2017}, Figure \ref{fig:b7images} suggests that the several
870\micron\ dust continuum sources have an adjacent companion of
smaller object. 
The detection threshold lower than $4.6\sigma$ results in selecting
sources without any counterpart in optical and NIR data, and thus 
they are likely to be spurious sources. This suggests that the
threshold that we apply is reasonable. As another check of the
reliability of the extracted sources, we also apply the same threshold
to the inverted data to select pixels with a large negative value and
then find only one negative detection. Therefore, we conclude that all
of the detections are real (Table \ref{tbl:B7sources}).

The flux densities of the dust continuum emission are measured with a
0.44''(=22 pixels)-diameter aperture, i.e., $\sim2.4\times$ the beam
size. The error of the flux densities are estimated from the 10,000
measurements with the same aperture at the random positions over the
individual images. The 1$\sigma$ error is derived by fitting a
Gaussian to the histogram of the random measurements.  
We also measure the flux densities in the uv-tapered map with the
synthesized beam size of 0.48$''$ $\times$ 0.46$''$ to verify whether
there is a flux of an extended component resolved out. The photometry
is performed in the same manner as in the natural weighting map, but
for the 1.0''-diameter aperture being used for the measurement in the
tapered map. In the case that the measurement in the natural weighting
map is consistent with that in the tapered map within the 1$\sigma$
error, we use the flux densities measured in the natural weighting
map. Otherwise, we use the measurement in the uv-tapered map.
The flux densities measured are shown in Table \ref{tbl:B7sources}. 

The 9 detections with $S_{\nu,870\micron}>$ 0.49 mJy in 0.61
arcmin$^2$ suggests that the number density of the 870\micron\ sources
in this region is a factor of 2--3 larger than expected from the
cumulative number counts of ALMA dust continuum sources in deep
general fields \citep[e.g.,][]{Fujimoto2016,Hatsukade2016}. However,
since the patchy coverage of the ALMA Band 7 data does not cover the
region where there are many cluster members, the number density would
be a lower limit. Indeed, \citet{Stach2017} report that the center of
this cluster is a $\sim7\times$ overdensity of 1.25mm dust continuum
sources.

Seven out of the nine sources have a counterpart of CO(2--1) emitters
detected in the Band 3 data (Figure \ref{fig:b7images}). Comparing the
intensity map of CO(2--1) with the map of Band 7, the position of both
dust continuum and CO emission coincides well. Also, the dust
continuum emission comes from the compact region of the center of the 
individual galaxy, which is similar to the previous studies for
high-$z$ galaxies \citep{Simpson2015,Ikarashi2015,Barro2016,Hodge2016,Tadaki2017a,Chen2017}. 
Although one source, B7.05, is located out of the Band 3 data
coverage, it has a counterpart of \oii\ emitters selected by
Subaru/Suprime-Cam narrowband imaging \citep{Hayashi2014}. However,
the source B7.04 is likely to be a foreground galaxy judging from the
appearance in the optical and NIR data 
\citep[see also][which show that this is a galaxy at $z=1.30$]{Stach2017}. 
Therefore, we regard the eight 870\micron\ sources as the cluster
member galaxies. Combining the results of \citet{Hayashi2017}, we have
found 18 detections in total in the ALMA Band 3 and Band 7 data (see
Figure \ref{fig:map} and Table \ref{tbl:ALMAsources}).    

\begin{center}
\begin{deluxetable*}{lcccccl}
 \tablecaption{Properties of galaxies detected in ALMA Band 7 \label{tbl:B7sources}}
 \tablehead{
  \multicolumn{1}{l}{ID} & \colhead{R.A.} & \colhead{Decl.} & \colhead{S/N\tablenotemark{a}} & \colhead{$S_{\nu, 870\micron}$} & \colhead{Detection\tablenotemark{b}} & \multicolumn{1}{l}{Counterpart}\\
  \colhead{} & \colhead{(J2000)} & \colhead{(J2000)} & \colhead{} & \colhead{(mJy)} & \colhead{in 1.25mm} & \colhead{}
 }
 \startdata
 ALMA.B7.01& 22 15 58.53& -17 37 47.6& 17.4& 2.58 $\pm$ 0.23\tablenotemark{c}& $\circ$ (3)& NB921 \oii\\
 ALMA.B7.02& 22 15 57.24& -17 37 53.4& 10.2& 1.08 $\pm$ 0.13\tablenotemark{\phn}& $\circ$ (6)& NB912+NB921 \oii\\
 ALMA.B7.03& 22 15 57.72& -17 37 45.2& 10.0& 0.71 $\pm$ 0.13\tablenotemark{\phn}& \nodata& NB912+NB921 \oii\\
 ALMA.B7.04& 22 15 59.97& -17 37 50.6& \phn7.4& 1.27 $\pm$ 0.18\tablenotemark{\phn}& $\circ$ (4)& foregrond galaxy ($z=1.30$)\tablenotemark{d}\\
 ALMA.B7.05& 22 16 05.40& -17 38 16.5& \phn7.1& 0.49 $\pm$ 0.14\tablenotemark{\phn}& ---& NB912+NB921 \oii\\
 ALMA.B7.06& 22 15 57.29& -17 37 58.0& \phn6.1& 1.12 $\pm$ 0.23\tablenotemark{c}& $\circ$ (7)& sBzK\\
 ALMA.B7.07& 22 15 58.77& -17 37 40.8& \phn4.8& 0.70 $\pm$ 0.20\tablenotemark{\phn}& $\circ$ (1)& NB912+NB921 \oii\\
 ALMA.B7.08& 22 16 00.38& -17 38 57.9& \phn4.7& 0.51 $\pm$ 0.14\tablenotemark{\phn}& ---& NB912+NB921 \oii\\
 ALMA.B7.09& 22 16 00.40& -17 37 50.8& \phn4.7& 1.05 $\pm$ 0.24\tablenotemark{c}& $\circ$ (5)& NB912 \oii
 \enddata
 \tablenotetext{a}{The signal-to-noise ratio in source detection.}
 \tablenotetext{b}{The numbers within parentheses show ID of the 1.25mm sources detected with ALMA Band 6 data by \citet{Stach2017}. The ``\nodata" means non-detection in the data, while the ``\phn---\phn" means that the data are not available for the galaxy.}
 \tablenotetext{c}{The uv-tapered map is used for the measurement of flux density. See the text for the details.}
 \tablenotetext{d}{The redshift is from \citet{Stach2017}.}
\end{deluxetable*}
\end{center}

\begin{center}
\begin{deluxetable*}{lcccccccccc}
 \tablecaption{Properties of the 18 galaxies detected in ALMA Band 3 and Band 7 \label{tbl:ALMAsources}}
 \tablewidth{\textwidth}
 \tablehead{
  \multicolumn{3}{l}{ID\tablenotemark{a}}& Redshift\tablenotemark{b}& Star-forming& M$_{\rm stellar}$& SFR$_{\rm SED-fit}$& $f_{\nu, 24\mu m}$& SFR$_{\rm UV+24\mu m}$& M$_{\rm gas, CO}$& M$_{\rm gas, dust}$\\
  \colhead{}& \colhead{(B3)}& \colhead{(B7)}& \colhead{}& \colhead{or Quiescent\tablenotemark{c}}& \colhead{(10$^{10}$\Msun)}& \colhead{(\Msun~yr$^{-1}$)}& \colhead{($\mu$Jy)}& \colhead{(\Msun~yr$^{-1}$)}& \colhead{(10$^{10}$\Msun)}& \colhead{(10$^{10}$\Msun)}
 }
 \startdata
ALMA.01 &01 &---     & 1.466& S& \phn8.13$^{+0.78}_{-0.54}$& \phn35$^{+35}_{- 2}$& 125 $\pm$ 10& \phn91$^{+ 7}_{- 8}$& 10.5$^{+0.5}_{-0.7}$& ---     \\
ALMA.02 &02 &\nodata& 1.450& S& \phn3.39$^{+0.69}_{-0.57}$& \phn31$^{+15}_{-13}$& \nodata& \nodata& \phn2.7$^{+0.6}_{-0.6}$& $<$3.7\\
ALMA.03 &03 &01      & 1.453& S& 11.22$^{+0.26}_{-1.22}$& \phn25$^{+21}_{- 6}$& \nodata& \nodata& 10.7$^{+0.8}_{-0.5}$& 13.8$^{+1.0}_{-1.2}$\\
ALMA.04 &04 &---     & 1.466& S& \phn3.89$^{+1.01}_{-0.65}$& \phn\phn6$^{+2.8}_{-2.9}$& \nodata& \nodata& \phn3.7$^{+0.6}_{-0.7}$& ---     \\
ALMA.05 &05 &---     & 1.467& S& \phn2.29$^{+0.66}_{-0.39}$& \phn48$^{+26}_{-20}$& \nodata& \nodata& \phn3.1$^{+0.6}_{-0.6}$& ---     \\
ALMA.06 &06 &---     & 1.467& S& 12.02$^{+2.43}_{-0.80}$& 145$^{+69}_{-32}$& 180 $\pm$ 11& 129$^{+ 9}_{- 6}$& 10.5$^{+0.7}_{-0.5}$& ---     \\
ALMA.07 &07 &06      & 1.452& S& \phn8.13$^{+0.19}_{-0.72}$& \phn35$^{+49}_{- 0}$& \nodata& \nodata& \phn5.8$^{+0.6}_{-0.6}$& \phn6.2$^{+1.2}_{-1.3}$\\
ALMA.08 &08 &---     & 1.457& S& \phn5.75$^{+0.27}_{-0.63}$& 105$^{+30}_{-34}$& \phn88 $\pm$ 10& \phn65$^{+ 8}_{- 7}$& \phn6.8$^{+0.5}_{-0.6}$& ---     \\
ALMA.09 &09 &03      & 1.468& S& 10.72$^{+0.25}_{-1.17}$& \phn47$^{+27}_{-24}$& \nodata& \nodata& \phn3.5$^{+0.5}_{-0.5}$& \phn3.8$^{+0.7}_{-0.7}$\\
ALMA.10 &10 &02      & 1.454& S& \phn3.98$^{+0.28}_{-0.43}$& \phn72$^{+28}_{-12}$& 125 $\pm$ 10& \phn91$^{+ 7}_{- 8}$& \phn8.1$^{+0.6}_{-0.5}$& \phn6.8$^{+0.8}_{-0.9}$\\
ALMA.11 &11 &07      & 1.451& S& \phn1.82$^{+0.58}_{-0.75}$& \phn17$^{+29}_{- 5}$& \nodata& \nodata& \phn6.6$^{+0.8}_{-0.9}$& \phn5.1$^{+1.5}_{-1.4}$\\
ALMA.12 &12 &\nodata& 1.445& S& \phn1.48$^{+0.22}_{-0.10}$& \phn54$^{+15}_{-11}$& \phn71 $\pm$ 10& \phn58$^{+ 7}_{- 6}$& \phn4.1$^{+0.6}_{-0.5}$& $<$4.6\\
ALMA.13 &13 &---     & 1.471& S& \phn6.03$^{+0.43}_{-2.71}$& \phn21$^{+55}_{- 3}$& \phn60 $\pm$ 10& \phn44$^{+ 8}_{- 6}$& \phn5.8$^{+0.7}_{-0.6}$& ---     \\
ALMA.14 &14 &09      & 1.451& Q& \phn9.12$^{+0.00}_{-1.88}$& \phn\phn3$^{+0.1}_{-2.5}$& \nodata& \nodata& \phn3.2$^{+0.6}_{-0.6}$& \phn5.8$^{+1.3}_{-1.3}$\\
ALMA.15 &15 &---     & 1.465& S& \phn3.63$^{+1.74}_{-0.24}$& \phn28$^{+10}_{-18}$& \phn85 $\pm$ 11& \phn62$^{+ 8}_{- 9}$& \phn6.6$^{+0.6}_{-0.9}$& ---     \\
ALMA.16 &16 &---     & 1.465& S& \phn3.09$^{+0.30}_{-0.52}$& \phn37$^{+29}_{-11}$& \nodata& \nodata& \phn8.5$^{+0.6}_{-0.9}$& ---     \\
ALMA.17 &17 &08      & 1.460& S& \phn2.45$^{+1.81}_{-0.46}$& 123$^{+51}_{-80}$& \nodata& \nodata& \phn5.8$^{+0.7}_{-0.6}$& \phn3.5$^{+1.0}_{-1.0}$\\
ALMA.18 &---&05      & 1.465& S& \phn2.51$^{+0.80}_{-0.11}$& \phn72$^{+15}_{-35}$& \nodata& \nodata& ---     & \phn3.4$^{+1.0}_{-0.9}$
 \enddata
 \tablenotetext{a}{The ``\nodata" means non-detection in the data, while the ``\phn---\phn" means that the data are not available for the galaxy (See Figure~\ref{fig:map}).}
 \tablenotetext{b}{The redshifts are derived from the CO emission lines for all but ALMA.18 \citep{Hayashi2017}. The redshift of ALMA.18 is from \citet{Hilton2010}.}
 \tablenotetext{c}{The galaxies are classified as star-forming (S) or quiescent (Q) galaxies based on the rest-frame U-V and V-J colors (Figure~\ref{fig:UVJ}).}
\end{deluxetable*}
\end{center}


\subsection{Archival imaging data}
\label{sec:ArchiveData}

We have optical images taken with Subaru/Suprime-Cam in 
$B, R_c, i', z'$, NB912 and NB921, which are used in our previous
studies \citep[e.g.,][]{Hayashi2010,Hayashi2014}. Other imaging data
in optical to infrared (IR) wavelengths are retrieved from public
archive.  
The Canada-France-Hawaii Telescope Legacy Deep Survey (CFHTLS-Deep)
provides us with the complementary optical images taken with
CFHT/MegaCam in $u*$ and $g'$, and the WIRCam Deep Survey
\citep[WIRDS,][]{Bielby2012} provides us with the NIR images taken
with CFHT/WIRCam in $J, H$ and $K_s$, all of which are retrieved from
the CFHT Science Archive.   
Other NIR images taken with a Wide Field Camera 3 (WFC3) on HST in
F125W, F140W, and F160W filters \citep{Beifiori2017} are also
retrieved from the HST archive.    
MIR data of Spitzer/IRAC 3.6--5.8 \micron\ and MIPS 24
\micron\ \citep{Hilton2010} are retrieved from the Spitzer Heritage 
Archive (SHA). We do not use IRAC 8.0 \micron\ data because the data
is not deep and thus many cluster members seem not to be detected in
8.0 \micron. 

A coadd image in MIPS 24 \micron\ is created by ourselves from the
Basic Calibrated Data (BCD) products retrieved from the archive. This
is because some pixels suffer from soft saturation in the individual
BCD products (see \citet{Hilton2010} for the details) and thus the
image quality at the north side is not good in the MIPS
24 \micron\ image reduced by the standard pipeline which can be
retrieved from the archive. Therefore, we mask the regions suffering
from soft saturation in the individual frames and then coadd them with
{\tt MOPEX} (MOsaicker and Point source EXtractor). Also, the pixel
scale is set to be 1.25 arcsec per pixel.   

\subsection{Catalogs of cluster member galaxies}
\label{sec:members}

In addition to the sample of the 18 ALMA sources, we have other
catalogs of cluster member galaxies selected from the previous studies
\citep{Hilton2010,Beifiori2017,Hayashi2010,Hayashi2011,Hayashi2014}. The
catalogs consist of \oii\ emission-line galaxies selected from imaging
with two narrowband filters \citep{Hayashi2010,Hayashi2014} and
galaxies confirmed by optical and NIR follow-up spectroscopy
\citep{Hilton2010,Hayashi2011,Hayashi2014,Beifiori2017}. The catalogs
provide us with additional 47 member galaxies within a radius of 
$1.1\times$ \R200\ (Figure \ref{fig:map}), where there are 31
\oii\ emission-line galaxies and 32 spectroscopically confirmed
galaxies. Note that the 16 galaxies in the sample of \oii\ emitters
overlap with those in the sample of spectroscopically confirmed
galaxies. Therefore, we can use the sample of 65 cluster member
galaxies within a cluster-centric radius of $\sim$ \R200 in this
study. The spatial distribution of the member galaxies is shown in
Figure~\ref{fig:map}. 

\subsubsection{multi-band photometry}

Photometry in the optical, NIR and MIR data is conducted for the ALMA
sources as well as the other cluster member galaxies. 
{\tt SExtractor} \citep{Bertin1996} is used for photometry in the
optical, NIR  and IRAC data. Since the seeing of the Subaru images is
1.09 arcsec which is worse than the other images \citep{Hayashi2014},
we match the point spread function (PSF) between the optical and NIR
images. Note that we do not match the PSF of the IRAC images. This is
because the PSF of IRAC data is quite different from the optical or
NIR data and thus it is not advisable to match the PSF of optical and
NIR data to that of IRAC data. We run {\tt SExtractor} in double
image mode. The $H$-band images with better seeing before the PSF is
matched, i.e., WFC3/F160W if available, otherwise WIRCam/$H$, is used
as a detection image. For the photometry in the optical and NIR data,
we measure magnitudes of the galaxies with a 2''-diameter aperture and
correct them for the aperture correction by 0.43 mag which is
estimated from growth curve of a PSF to derive total magnitudes. 
The magnitudes are also corrected for the Galactic absorption assuming
the extinction law of \citet{Cardelli1989}. 
For the photometry in the IRAC data, magnitudes are measured with a
3''-diameter aperture, and then we apply the aperture correction of
0.54, 0.63, and 0.83 mag in [3.6], [4.5], and [5.8] to estimate the
total magnitude.    

We conduct PSF-fitted photometry with {\tt IRAF/DAOPHOT} in the MIPS
data, according to the previous studies \citep[e.g.,][]{Magnelli2009}. 
We use the IRAC 3.6 \micron\ data as a reference image. We fit the PSF
to the MIPS data in each position of IRAC 3.6 \micron\ sources, and
measure the flux in 24 \micron. We check the residual image to make
sure that the fitting works well. We apply the aperture correction of
0.53 mag which are estimated from growth curve of a PSF. To combine
the photometry in the MIPS data with that in the shorter wavelengths,
an aperture with a 1.5'' radius is used for the source
matching. Seven of the 18 ALMA sources are detected in 24 \micron\ 
(Table~\ref{tbl:ALMAsources}).

\subsubsection{stellar mass, SFR, and rest-frame colors}

\begin{figure}
  \begin{center}
    \includegraphics[width=0.47\textwidth, clip, trim=10 15 10 10]{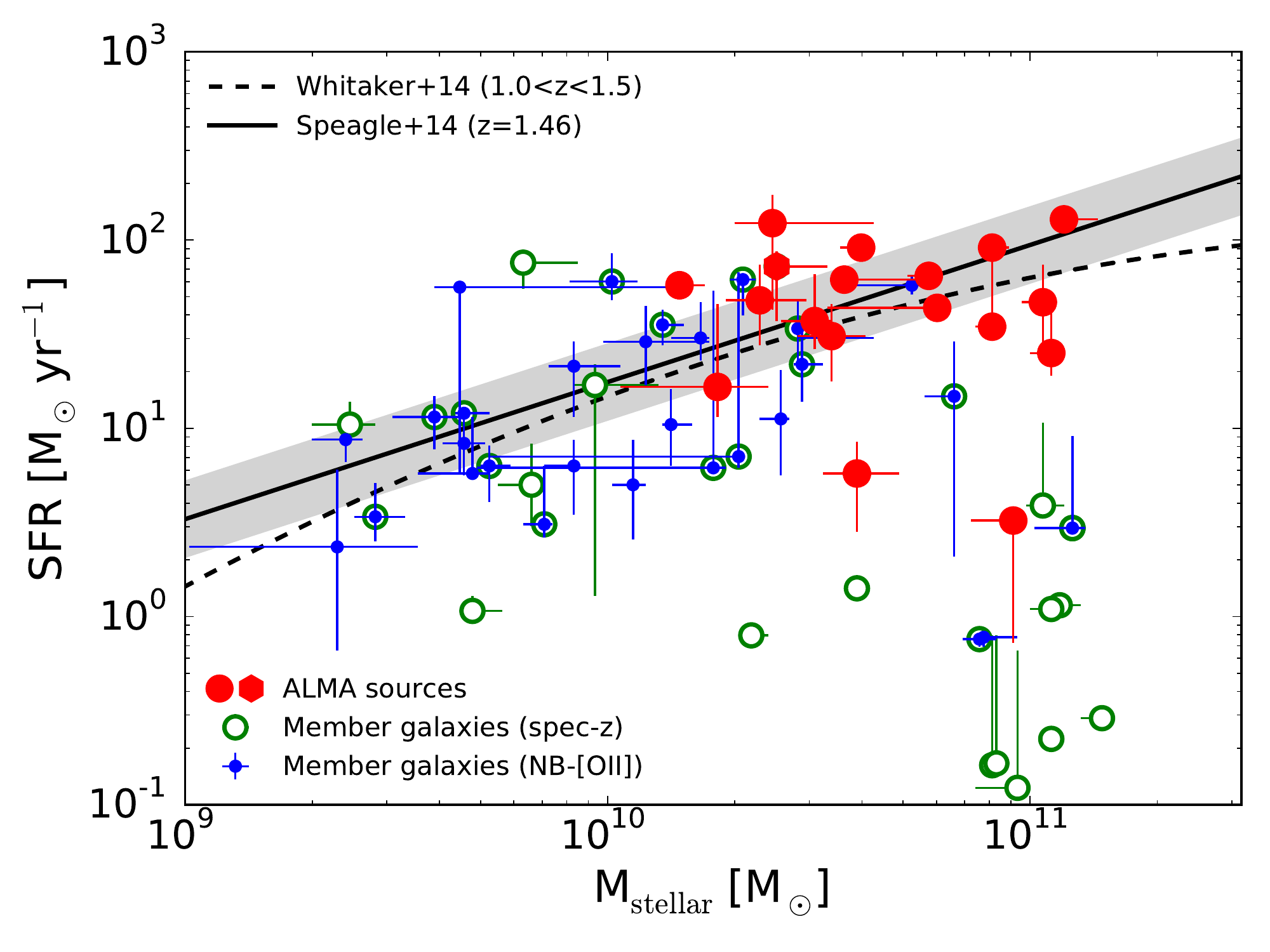}    
    \caption{ 
      SFRs as a function of stellar mass for the cluster member
      galaxies within a radius of $\sim$ \R200. The red symbols show
      the 18 ALMA sources: 17 CO(2--1) emitters are shown by circle,
      while the dust continuum source is shown by hexagon. 
      The spectroscopically confirmed galaxies are shown by green
      circles and the \oii\ emitters are shown by blue circles. 
      The solid line with the gray region shows the main sequence with
      $\pm$0.2 dex given by \citet{Speagle2014}, and the dashed line
      shows the one by \citet{Whitaker2014}.      
      \label{fig:MS}
    }
  \end{center}  
\end{figure}

The multi-band photometry covering the optical to MIR wavelengths is
used to estimate stellar masses and SFRs of the galaxies. We use the
C++ version\footnote{\url{https://github.com/cschreib/fastpp}} of 
{\tt FAST} code \citep{Kriek2009} to perform the spectral energy
distribution (SED) fitting. The redshifts of the galaxies are fixed at
ones estimated from CO(2--1) lines, optical and NIR spectroscopy, and
narrowband response functions, where the redshifts determined from the
former have higher priority when the redshifts from several methods
are available. The model SED templates of galaxies are generated by
the code of \citet{BC03}. 
Star formation histories of the exponentially declining model are
adopted, where we set an e-folding time of $\log(\tau/{\rm yr})=$ 
8.5--10.0 with $\Delta\log(\tau/{\rm yr})=$ 0.1 \citep{Wuyts2011}. 
The ages of 0.1--10.0 Gyr are acceptable with a step of
$\Delta\log(\rm age/yr)=$ 0.1.
The extinction curve of \citet{Calzetti2000} is assumed and $A_V$
ranges from 0.0 to 3.0. Metallicity is fixed to the solar value.
Monte Carlo simulation is performed 100 times for each galaxy to
estimate a $1\sigma$ error in stellar mass and SFR.  

The intrinsic SFRs estimated are sensitive to reliability of
correction for dust extinction. MIR data are useful to estimate a
component of SFR obscured by dust, which suggests that it is not easy
to estimate the dust obscured SFR from the rest-frame UV and optical 
data \citep{Tadaki2017a,Whitaker2017}. Several studies in a galaxy
cluster at $z\sim0.4$ suggest that galaxies in higher density regions
tend to be more dusty \citep{Koyama2013b,Sobral2016}. 
Therefore, if a galaxy has a detection in 24 \micron, we estimate the
SFR from the combination of UV and IR luminosities. Otherwise, we use
the dust-corrected SFR derived from the SED fitting described
above. The IR luminosities are estimated from 24 \micron\ fluxes using 
a conversion factor given by \citet{Wuyts2008}, and the UV
luminosities are estimated from the rest-frame 2800 \AA\ luminosity of
the best-fit SED. Then, using the equation given by \citet{Wuyts2011},
\begin{equation}
\frac{\rm SFR_{UV+IR}}{M_\odot\ {\rm yr^{-1}}} = 1.09\cdot10^{-10}\cdot\frac{L_{IR}+3.3L_{2800}}{L_\odot},
\end{equation}
the UV+IR luminosities are converted to SFR$_{\rm UV+IR}$.

Table \ref{tbl:ALMAsources} lists the stellar masses and SFRs for the
ALMA sources. Figure~\ref{fig:MS} shows the SFRs of the galaxies as a
function of stellar mass. We also plot the main sequence at redshift
of $z\sim1.46$ from the literature \citep{Whitaker2014,Speagle2014}. 
Most of the galaxies detected in ALMA data are located on or above the
main sequence (MS) at the redshift (i.e., within $\pm0.2$ dex or
higher from the MS). The other ALMA sources below the main sequence
are massive galaxies with $\ga10^{10.6}$ \Msun. 

\begin{figure}
  \begin{center}
    \includegraphics[width=0.55\textwidth, clip, trim=40 0 0 0]{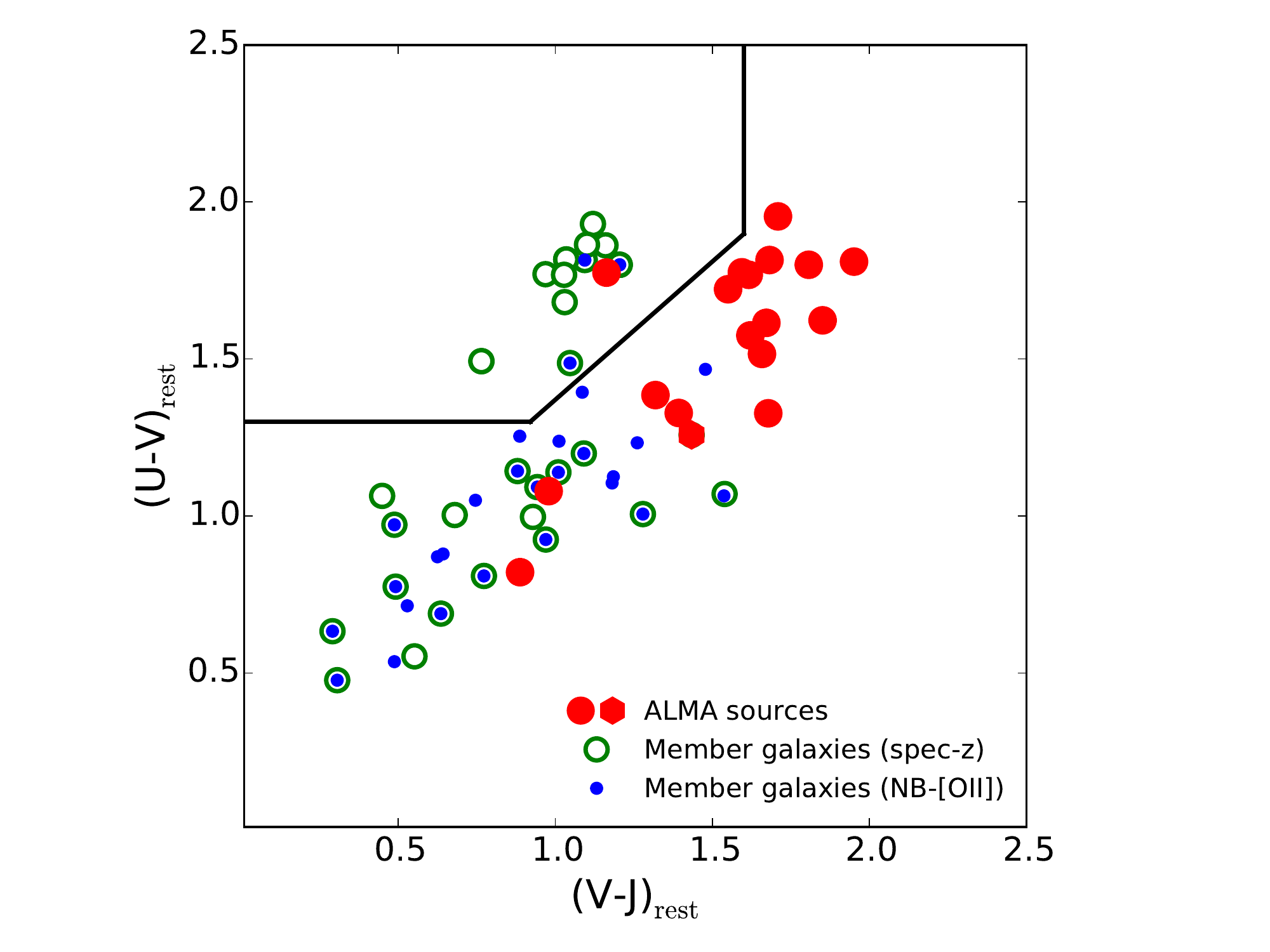}    
    \caption{ 
      U-V versus V-J colors in the rest-frame. The symbols are the
      same as Figure~\ref{fig:MS}. The solid line is a boundary to
      distinguish quiescent galaxies from star-forming galaxies
      \citep{Williams2009}. 
      \label{fig:UVJ}
    }
  \end{center}  
\end{figure}

Figure~\ref{fig:UVJ} shows the rest-frame U-V versus V-J colors of the
member galaxies. The U-V and V-J colors are derived from the best-fit
SED. According to \citet{Brammer2011}, we use the response function of
U and V filters defined by \citet{MaizApellaniz2006} and 2MASS-J
filter to calculate the colors. The UVJ diagram is widely used to
distinguish quiescent galaxies from star-forming galaxies
\citep[e.g.,][]{Labbe2005,Williams2009}. Almost all of the CO(2--1)
lines and dust continuum emissions are detected from star-forming
galaxies. Along the sequence of the star-forming galaxies in the UVJ
diagram, the ALMA sources tend to have redder colors, implying that
the CO line and dust continuum emission are easier to be detected from
more dusty star-forming galaxies \citep[e.g.,][]{Tadaki2015,Rudnick2017}. 
On the other hand, few quiescent member galaxies have either CO line
or dust continuum emission detected. Indeed, the ALMA.14 is only
classified as a quiescent galaxy (see also Table~\ref{tbl:ALMAsources}).    

\section{Results}
\label{sec:results}

\subsection{molecular gas mass}
\label{sec:results.Mgas}

We estimate molecular gas mass from luminosities of CO(2--1) emission
line and dust continuum emission for the 18 cluster member galaxies
with detection in ALMA data. The CO(2--1) luminosities are derived in
\citet{Hayashi2017} from the intensity map integrated in velocity by
the width of emission line (2$\times$FWHM). 
The luminosity, $L'_{\rm CO(2-1)}$, ranges (4.5--22) $\times10^{9}$ K
km $^{-1}$ pc$^2$. The conversion factor from the CO luminosity to the
molecular gas mass given by \citet{Tacconi2018} is adopted to
estimate the molecular gas mass. 
We use $L'_{\rm CO(1-0)}/L'_{\rm CO(2-1)}=1.2$ and $\alpha_{\rm CO(1-0)}=4.36$. 
Note that the conversion factor is corrected for the metallicity
dependence through the stellar mass -- metallicity relation
\citep{Genzel2012,Bolatto2013} and thus the conversion factors that we 
adopt range from 4.99 to 7.31 which are dependent on
the stellar mass. We also use the equation (16) of
\citet{Scoville2016} to estimate molecular gas mass from dust
continuum emission at 870 \micron, where the dust temperature of
$T_d=25$ K is assumed and the metallicity (i.e., stellar
mass)-depended ratio of molecular gas to dust mass is taken into
account according to \citet{Tacconi2018}.  

\begin{figure}
  \begin{center}
    \includegraphics[width=0.45\textwidth, clip, trim=10 15 10 0]{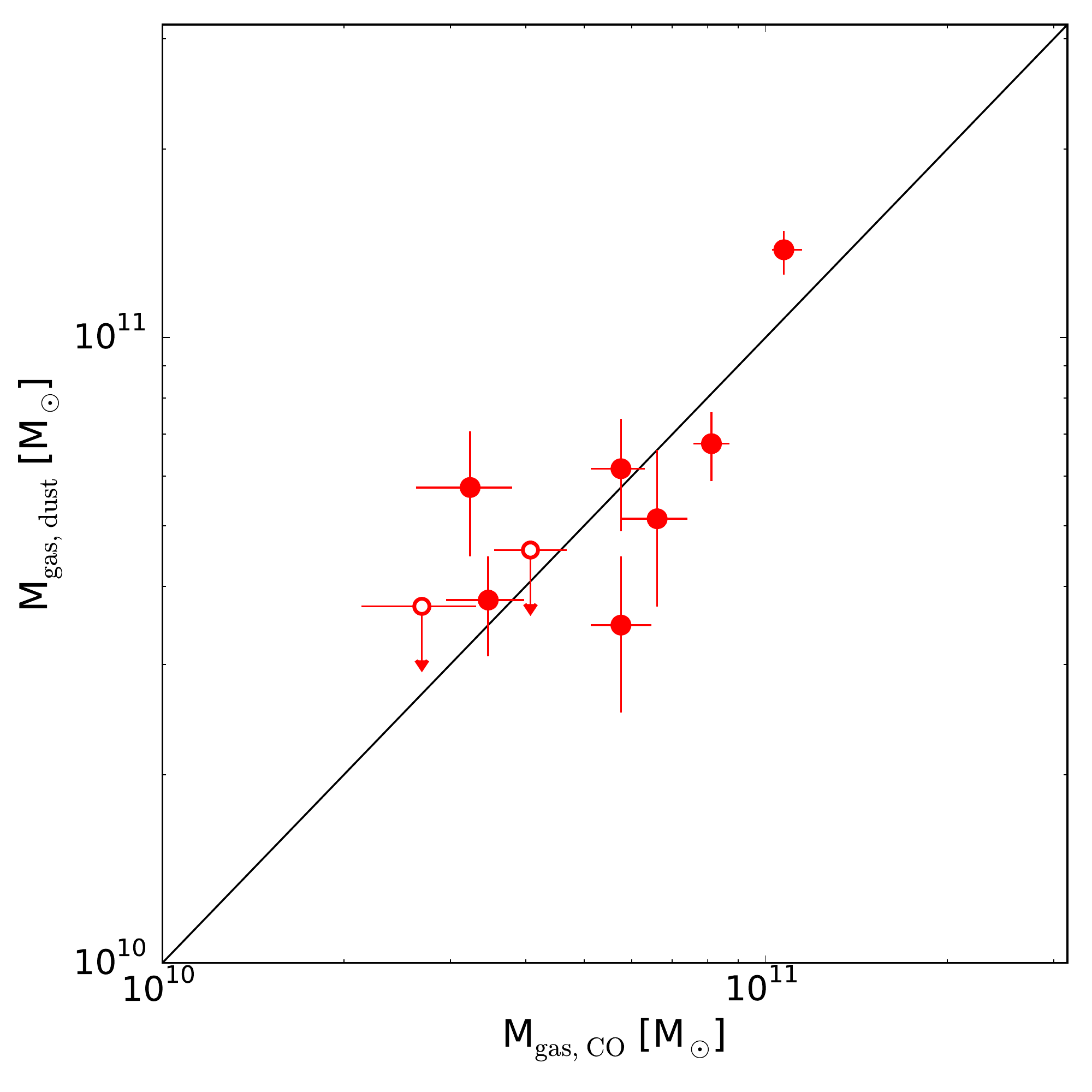}     
    \caption{
      Comparison between molecular gas estimated from CO(2--1) and
      that from dust continuum for galaxies in the area 
      where both Band 3 and Band 7 data are available. The filled
      circles show the cluster members with both CO(2--1) line and
      dust continuum detected. The open circles show the members with
      CO(2--1) lines detected but dust continuum not detected. The
      upper limits are estimated from the flux densities of 4.6$\sigma$
      noise level in the Band 7 data at the position of CO(2--1) line
      which is the same as the detection limit in \S~\ref{sec:data.band7}.
      \label{fig:MgasCO_MgasDust}}
  \end{center}  
\end{figure}

Figure~\ref{fig:MgasCO_MgasDust} compares the molecular gas mass from
CO(2--1) luminosity with that from dust continuum luminosity for 
galaxies with a detection in both CO and dust continuum. The molecular
gas masses estimated from the two ways are consistent. Among the
galaxies with CO(2--1) line in the area covered by ALMA Band 7 data,
two CO(2--1) emitters are not detected in Band 7 (The IDs are ALMA.02
and ALMA.12). The two galaxies have the lowest CO(2--1) luminosities
among the galaxies for which the Band 7 data are available. The
3.8$\sigma$ source is seen in the Band 7 data near the position of
ALMA.02, while no source at more than $3\sigma$ is seen around the
ALMA.12 within the synthesized beam of Band 3 data. We plot their
upper limit of the molecular gas mass from dust continuum in
Figure~\ref{fig:MgasCO_MgasDust}, suggesting that the gas mass from CO
is not discrepant with the upper limit from dust continuum for the two
galaxies. Hereafter, if the galaxies have CO luminosity available, we 
use the gas mass derived from CO. Otherwise, we use the gas mass
derived from dust continuum, namely we use  M$_{\rm gas,dust}$ for the
galaxy of ALMA.18 only. The molecular gas masses derived here
are listed in Table~\ref{tbl:ALMAsources}. Moreover, we estimate an
upper limit of the molecular gas mass from the upper limit of the CO
luminosity (\S~\ref{sec:data.band3}) for the individual member 
galaxies in the area covered by the Band 3 data.

The conversion factor applied to derive molecular gas mass from CO
luminosity is one of the major uncertainties in the measurement. It is
not obvious which conversion factors should be used. \citet{Stach2017}
argue that at least two member galaxies in this cluster prefer the
conversion factor $\alpha_{CO}=1$ based on the comparison between gas
mass from the CO luminosity and dynamical mass from the width of CO
line. However, we find that the gas masses we estimate by the
different ways, i.e, CO and dust continuum, are consistent with each
other even for the two galaxies (ALMA.07 and ALMA.10 in
Table~\ref{tbl:ALMAsources}). Moreover, we compare our results with
the scaling relation for field galaxies given by \citet{Tacconi2018}
in Section~\ref{sec:discussion}. The conversion factors that we apply
in this work are the same as in \citet{Tacconi2018}, which enables a
fair comparison of our results with the scaling relation.

\begin{figure}
  \begin{center}
    \includegraphics[width=0.47\textwidth, clip, trim=10 15 10 10]{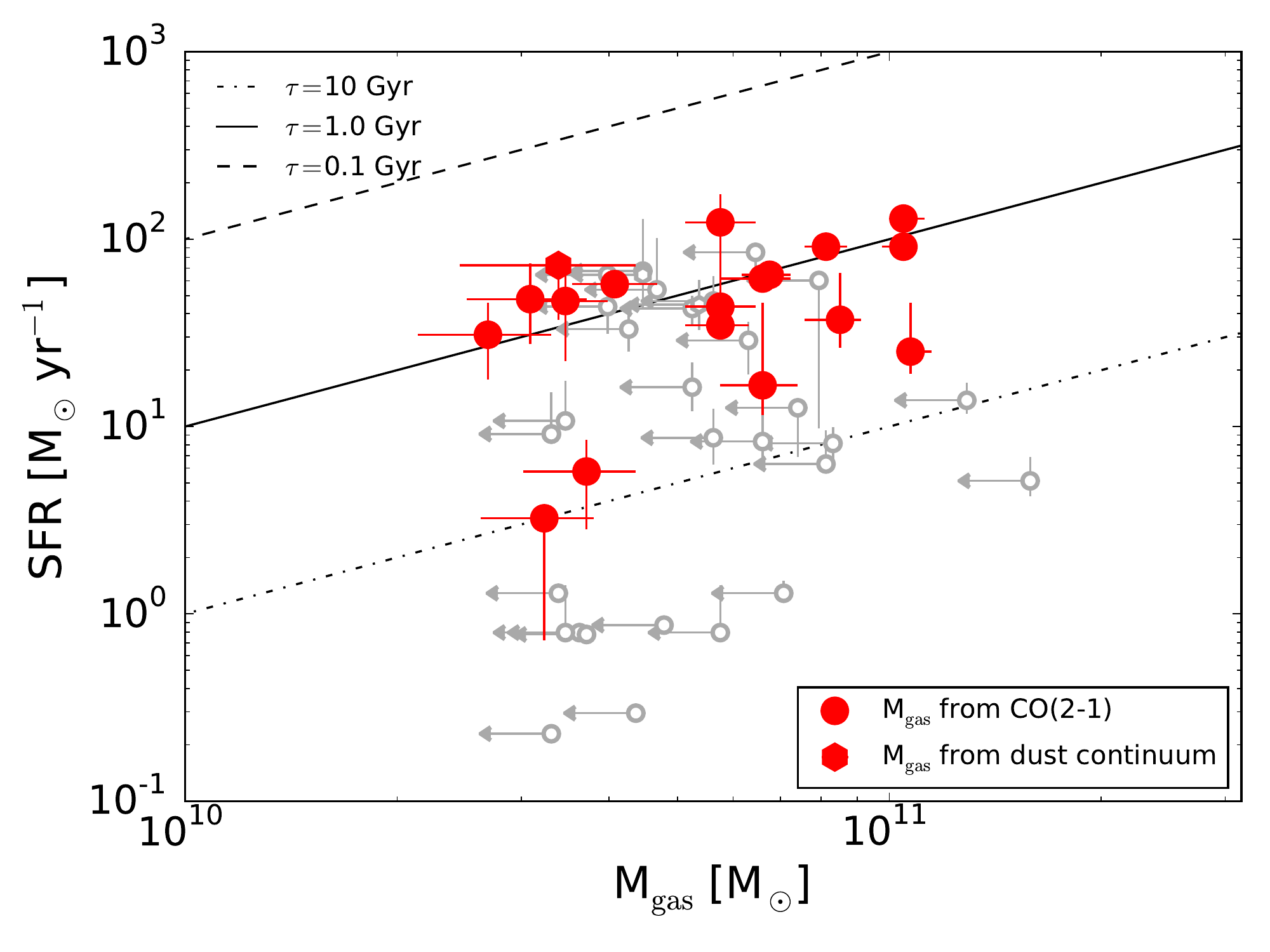}     
    \caption{
      SFRs as a function of molecular gas mass. The red circles are 17
      CO(2--1) emitters and the red hexagon is the dust continuum
      source. The gray symbols show the upper limit of molecular gas
      at the SFR estimated. 
      The dash-dotted, solid, and dashed lines show a constant
      depletion time scale, SFR/M$_{\rm gas}$, of 0.1, 1.0, and 10
      Gyr, respectively.  
      \label{fig:SFR_Mgas}}
  \end{center}  
\end{figure}

Figure~\ref{fig:SFR_Mgas} shows the SFRs of the member galaxies as a
function of molecular gas mass. There may be a mild trend that
galaxies with larger gas masses have larger SFRs, although larger
sample is required for confirmation of this trend. At a given
molecular gas, the galaxies can have a wide range of SFRs ($\sim$1.0
dex), indicating a wide range of star formation efficiency among the
cluster member galaxies. The PHIBSS survey shows larger SFRs with
$\sim0.5$ dex dispersion at a given molecular gas for field galaxies
at $z=$ 1.0--2.5 \citep{Tacconi2013}. These suggest that besides
molecular gas, other factors also have an impact on the star formation
activity of cluster galaxies.    

\begin{figure*}[t]
  \begin{center}
    \includegraphics[width=\textwidth, clip, trim=5 15 5 10]{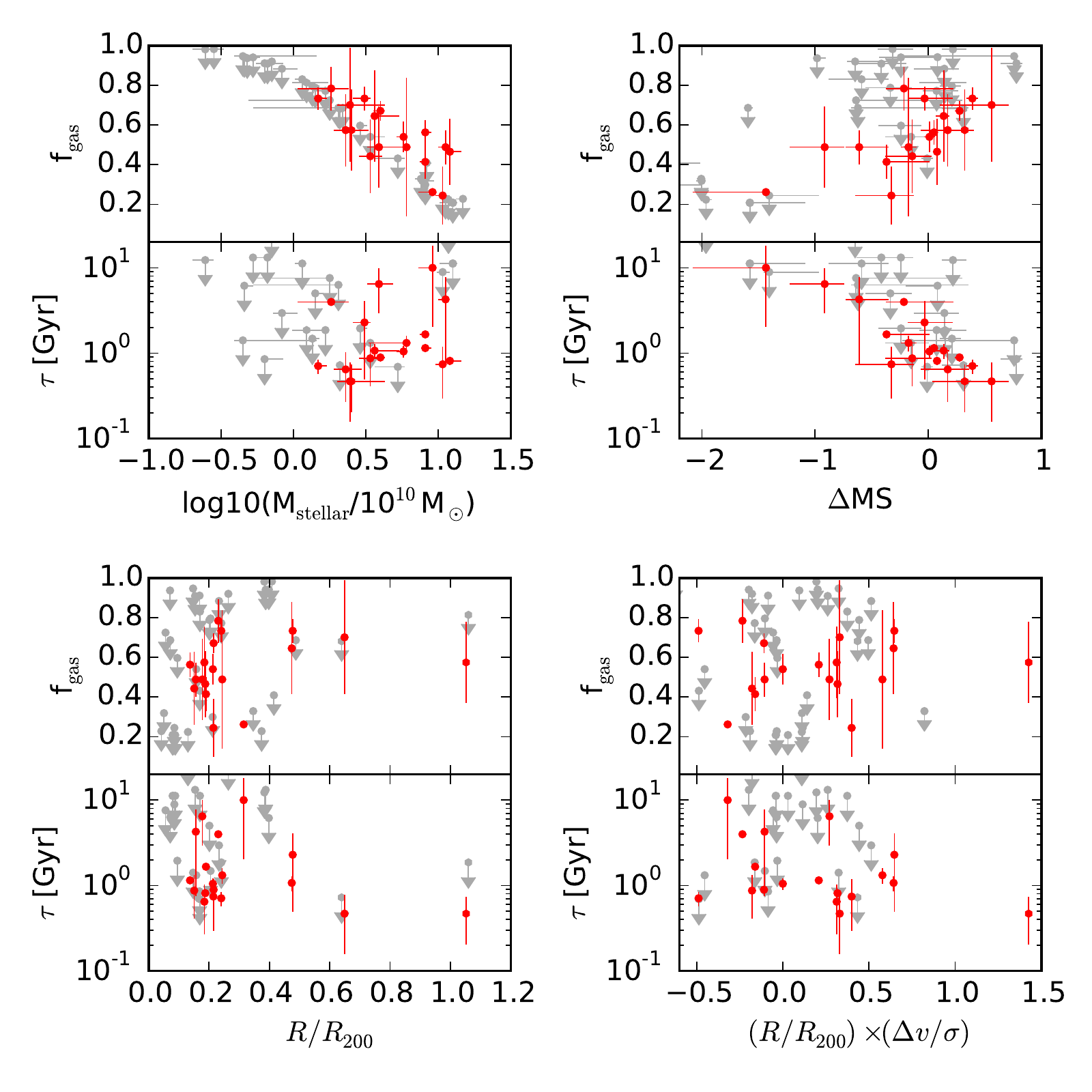}    
    \caption{
      Molecular gas fraction, $f_{\rm gas}=M_{\rm gas}/(M_{\rm gas}+M_{\rm stellar})$, 
      and depletion time, $\tau=M_{\rm gas}$/SFR, of the cluster
      member galaxies as a function of stellar mass (M$_{\rm stellar}$),
      offset from the main sequence ($\Delta$MS), cluster-centric
      radius ($R/R_{200}$), and accretion phase
      ($(R/R_{200})\times(\Delta v/\sigma)$), where we assume the MS
      of star-forming galaxies at $z=1.46$ given by \citet{Speagle2014}. 
      The red symbols show the 18 ALMA sources detected in CO(2--1) or
      dust continuum. The gray symbols show the upper limit of gas
      fraction and deletion time for the other member galaxies without
      detection in ALMA data. 
      \label{fig:fgas_Tdep}
    }
  \end{center}
\end{figure*}

\begin{figure*}
  \begin{center}
    \includegraphics[width=\textwidth, clip, trim=5 15 5 10]{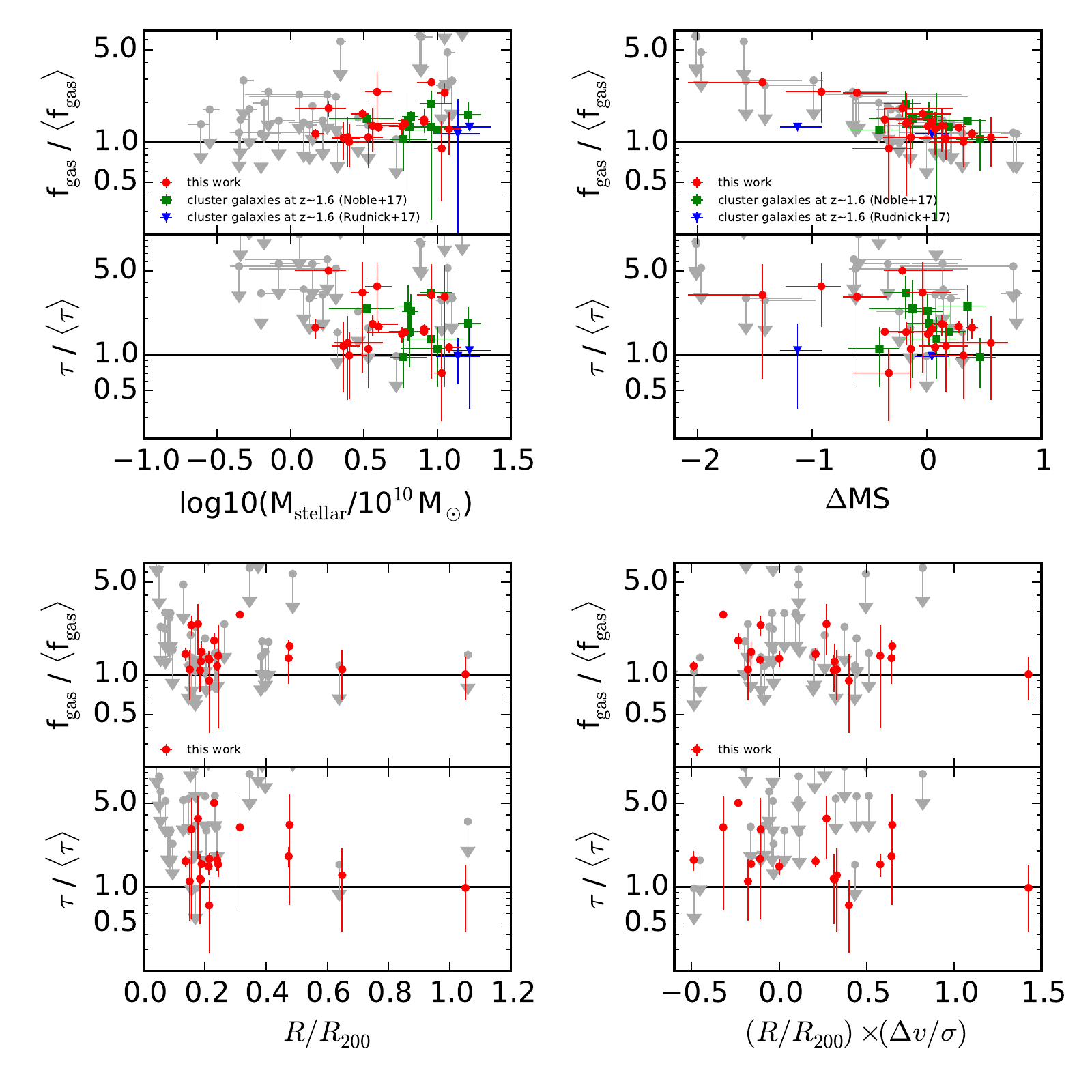}    
    \caption{Same as Figure~\ref{fig:fgas_Tdep}, but the gas fraction
      and the depletion time are compared with what is expected from
      the scaling relations for field galaxies given by
      \citet{Tacconi2018}. The main sequence is derived from the
      literature of \citet{Speagle2014}. 
      The scaling relations are functions of redshift, stellar mass,
      and ratio of specific SFR to that of MS galaxies with a given
      stellar mass at the redshift, and thus take account of the
      redshift evolution of specific SFR.
      The horizontal line in each panel shows the gas fraction and the
      depletion time on the scaling relations. The red circles show our
      results. The green squares show the results of \citet{Noble2017}
      for galaxies in three galaxy clusters at $z\sim1.6$ and the blue
      triangles show the results of \citet{Rudnick2017} for galaxies
      in a galaxy cluster at $z\sim1.6$.
      \label{fig:fgas_Tdep_SR}
    }
  \end{center}
\end{figure*}

\subsection{Gas mass fraction and depletion time scale}

Fraction of molecular gas mass to the sum of stellar and gas mass,
$f_{\rm gas}=M_{\rm gas}/(M_{\rm gas}+M_{\rm stellar})$, and 
depletion time scale, $\tau=M_{\rm gas}$/SFR, are useful to
characterize evolutionary phases of the galaxies. The fraction of
gas mass can imply how the member galaxies have gas reservoirs and then
can proceed to form stars, and the depletion time is a time scale
reflecting the efficiency of star formation in the galaxies. Since we
expect that star-forming galaxies located in the central region of the
cluster are good candidates of present-day massive early-type
galaxies, it is important to investigate the fraction of gas mass and
the depletion time scale for the member galaxies to discuss the
evolution of the cluster galaxies.  

We investigate how the gas fraction and the depletion time are related
with the evolution of cluster galaxies. We here focus on the following
four factors: stellar mass, offset from the MS of star-forming
galaxies, cluster-centric radius, and accretion phase based on
phase-space. The former two factors are related with the properties of
galaxies themselves. Stellar mass of galaxies is one of the most
important properties showing the tight correlation with other galaxy
properties such as star formation, metallicity, and size.  
The offset of the MS is also an important factor to discuss the
relation between the gas reservoirs and star formation activity in
cluster galaxies. On the other hand, the latter two factors should
give us insight into evolutionary processes peculiar to galaxy
clusters after they belong to the galaxy cluster. A phase-space
diagram is a useful tool to characterize the accretion state of
cluster member galaxies relatively free from effects due to the 2D
projected positions with respect to the cluster center
\citep{Noble2013,Noble2016,Jaffe2015,Muzzin2014}. \citet{Hayashi2017}
show that the CO emitters tend to be distributed at the edge of the
virialized region or in the region of relatively recent accretion. 
The galaxies with CO line detected disappear from the very center of
the cluster. They argue that the gas-rich galaxies with CO detections
have spent only relatively short times within the cluster.   

Figure~\ref{fig:fgas_Tdep} shows gas fraction ($f_{\rm gas}$) and
depletion time ($\tau$) of the 18 gas-rich member galaxies as a
function of stellar mass, offset from the MS, cluster-centric radius,
and accretion phase, respectively. We assume the MS of star-forming
galaxies at $z=1.46$ given by \citet{Speagle2014} which investigate
the evolution of MS up to $z\sim6$ by compiling 25 studies from the
literature. The offset of the MS is derived from a difference between
the SFR and the expectation from the MS at a given stellar mass. We
also plot the upper limits of gas fraction and depletion time for the
other member galaxies.  

The gas fractions in the massive galaxies with 
M$_{\rm stellar}\sim10^{11}$ \Msun\ are roughly less than half and the
depletion time scale is $\ga1$ Gyr. In particular, these galaxies
with $\Delta$MS $<-1$ show lower gas fraction of $<1/3$, suggesting that
massive quiescent galaxies no longer have large gas reservoir and
efficient star formation in the cluster center. 
On the other hand, the galaxies above the MS show larger gas fraction
and smaller depletion time scale as the offset from the MS is larger. 
The galaxies with larger offset from the MS tend to be gas-rich
galaxies forming stars in starburst phase. As long as we focus on the
galaxies with CO and dust continuum detected, most of them show 
$f_{\rm gas}\ga0.4$ and $\tau\ga1$ Gyr and there is no strong
dependence of gas fraction and depletion time on the cluster-centric
radius and the accretion phase. However, it is also a fact that
massive quiescent galaxies that show $f_{\rm gas}\la0.2$ and
$\tau\sim10$ Gyr exist in the cluster center. These are few member
galaxies with an intermediate gas fraction and depletion time
scale. Some rapid processes may be able to reduce the gas reservoirs
in cluster galaxies.

We integrate the molecular gas mass, stellar mass and SFR for the
member galaxies within a radius of $0.5R_{200}$ to estimate an average
gas fraction,  
$\langle f_{gas} \rangle=\sum{\rm M_{gas}}/\sum{\rm M_{stellar}}$, and an
average depletion time scale, 
$\langle \tau \rangle=\sum{\rm M_{gas}}/\sum{\rm SFR}$, in the cluster. 
The gas fraction and depletion time ranges 
$\langle f_{gas} \rangle=$0.29--0.53 and 
$\langle \tau \rangle=$0.83--2.3 Gyr, where the upper limit is derived
by taking account of the upper limit of gas mass for the member
galaxies without ALMA detection and the lower limit is derived by
assuming no gas mass for these member galaxies.

\section{Discussion}
\label{sec:discussion}

\subsection{Comparison with the scaling relations}
\label{sec:discussion.SR}

Scaling relations of gas to stellar mass ratio ($\rm M_{gas}/M_{stellar}$) 
and depletion time scale on both specific SFR and its offset from the
MS are derived by \citet{Tacconi2018} from the compilation of more
than one thousand measurements of molecular gas mass for galaxies at
$z=$0--4 with a wide range of stellar mass and SFR. We use the scaling
relations to compare our results with the representative populations
in general fields at similar redshifts, where we assume the MS of
\citet{Speagle2014} as in \citet{Tacconi2018}. 
Figure~\ref{fig:fgas_Tdep_SR} shows the ratio of molecular gas
fraction and depletion time to what is expected from the scaling
relations as a function of stellar mass, offset from the MS,
cluster-centric radius, and accretion phase based on phase space.  

The member galaxies with CO and/or dust continuum detected in the
cluster tend to have larger gas fraction and larger depletion time, 
compared with those from the scaling relations. 
The cluster galaxies with large offset below from the MS also have the
gas fraction larger than that for field galaxies, nevertheless they
are quenching star formation. Judging from the distribution of the
cluster galaxies in the SFR-$\rm M_{stellar}$ diagram
(Figure~\ref{fig:MS}), the larger depletion time scale is due not to
the lower SFR but to the larger amount of gas.  
The results may imply that the infalling regions and filaments around
galaxy clusters are easier to feed gas to member galaxies, which
results in the larger gas fraction in cluster galaxies. Also, some
environmental effects peculiar to galaxies associated with galaxy
clusters may reduce the efficiency of star formation. We speculate
that the shock-heating by ram pressure can be one of the causes of the
low efficiency 
\citep{Jachym2014,Wong2014}. Although the statistics is poor, it seems 
that the member galaxies in $R\ga$ 0.5\R200\ or with phases accreting
more recently have gas fraction and depletion time consistent with the
scaling relations. On the other hand, the member galaxies infalling to
closer to the cluster center can have larger gas fraction and larger
depletion time than the scaling relations. This supports that some
environmental effects have impacted the galaxies while moving within
the galaxy cluster. 
Moreover, if the negative feedback from active galactic nucleus (AGN)
and/or supernova (SN) works on the galaxies \citep{Carniani2017}, the
efficiency of star formation would be further reduced, although some
studies suggests the possibility of the positive feedback by AGNs
\citep{Kakkad2017}.

However, we cannot completely exclude the possibility that we overestimate the
molecular gas mass. Although we use a conversion factor, 
$\alpha_{\rm CO}$, dependent on stellar mass (i.e., metallicity
through the mass-metallicity relation), if the cluster galaxies have
higher metallicity at a given stellar mass, then the actual conversion
factor should be smaller than what we apply in this work. 
This can result in the overestimation of molecular mass by a factor of
$\sim1.5$, based on the conversion factors that we apply in
\S~\ref{sec:results.Mgas}. The mass-metallicity relation in galaxy
clusters at high redshifts is still controversial, however, several
studies suggest that less massive galaxies in high-$z$ galaxy clusters
tend to be more metal-rich than the field galaxies, while massive
galaxies do not show such a difference between galaxy clusters and
fields \citep{Kulas2013,Shimakawa2015a}.

\subsection{Comparison with other clusters at similar redshifts}
\label{sec:discussion.otherCLs}

To discuss how representative the results that we have found are in 
galaxy clusters at $z\sim1.5$, we compare with two studies in galaxy
clusters at $z\sim1.6$ \citep{Noble2017,Rudnick2017}.   
\citet{Noble2017} have detected CO (2-–1) emission lines from 11
gas-rich galaxies in three galaxy clusters at $z\sim1.6$ using
ALMA. They argue that the cluster galaxies tend to have enhanced gas
fractions compared with the field scaling relations at $z=1.6$ but
they have depletion timescales consistent with the field
galaxies. These three clusters are found by the Spitzer Adaptation of
the Red-Sequence Cluster Survey
\citep[SpARCS;][]{Wilson2009,Muzzin2009}. They estimate the cluster
mass of $\ga10^{14}$ \Msun\ from the richness of member galaxies 
\citep{Noble2017}. 
\citet{Rudnick2017} have detected CO(1-–0) emission lines from two
massive galaxies in a confirmed $z=1.62$ galaxy cluster
\citep{Papovich2010,Tanaka2010} using the Karl G. Jansky Very Large
Array (VLA). They argue that the gas fractions and star formation
efficiencies of the galaxies in the cluster are comparable to the
field galaxy scaling relations. The cluster mass is estimated to be
$1.1\times10^{14}$ \Msun\ from XMM-Newton X-ray data \citep{Papovich2010,Tanaka2010}. 
Because the cluster mass of the XMMXCS J2215.9-1738 cluster is
estimated to be $\sim3\times10^{14}$ \Msun, all of these clusters are
systems with similar mass scale. 

The molecular masses for the galaxies associated with the galaxy
clusters at $z\sim1.6$ are derived in the same manner as in Section
\ref{sec:results.Mgas} using the information available from the literature 
\citep{Noble2017,Rudnick2017} for proper comparison with our
results. We use the stellar masses and SFRs of the galaxies shown in
the literature. The comparison with the results in the galaxy clusters
at $z\sim1.6$ show that our results are consistent with the other
clusters at similar redshifts (Figure \ref{fig:fgas_Tdep_SR}). 
Therefore, we conclude that cluster galaxies at $z\sim1.5$ can have
the molecular gas fraction larger than what the field galaxies have. 
While the depletion time scale of the massive cluster galaxies with 
$\sim10^{11}$ \Msun\ are similar to the field galaxies, less massive
galaxies can have larger depletion time.

\begin{center}
\begin{deluxetable*}{lccccccccc}
 \tablecaption{Average properties of the confirmed member galaxies within a half of $R_{200}$ obtained from stacked spectra in Band 3 \label{tbl:stack}}
 \tablewidth{\textwidth}
 \tablehead{
 & \multicolumn{2}{c}{Number\tablenotemark{a}}& $\rm \langle M_{stellar}\rangle$\tablenotemark{b}& $\rm \langle SFR \rangle$\tablenotemark{b}& $S\Delta v$\tablenotemark{c}& M$_{\rm gas, CO}$& f$_{\rm gas}$& $\tau$\\
  \colhead{}& \colhead{}& \colhead{}& \colhead{(10$^{10}$\Msun)}& \colhead{(\Msun~yr$^{-1}$)}& \colhead{(Jy km s$^{-1}$)}& \colhead{(10$^{10}$\Msun)}& \colhead{}& \colhead{(Gyr)}
 }
 \startdata
Quiescent& 12& (\phn1)& 11.0& \phn1& $<$0.07& $<$0.93& $<$0.08& $<$9.78\\
Star-forming& 27& (15)& \phn2.3& 31& \phm{$<$}0.21& \phm{$<$}3.83& \phm{$<$}0.63& \phm{$<$}1.24\\
Star-forming w/o CO& 12& (\phn0)& \phn0.6& \phn7& $<$0.04& $<$1.05& $<$0.62& $<$1.58
 \enddata
 \tablenotetext{a}{The number of galaxies stacked. The values within parentheses are the number of galaxies with CO(2--1) detected individually.}
 \tablenotetext{b}{The median values in the samples.}
 \tablenotetext{c}{Estimated from the intensity map integrated in velocity by 400 km s$^{-1}$.}
\end{deluxetable*}
\end{center}


\subsection{Molecular gas reservoirs of quiescent galaxies}
\label{sec:discussion.stack}

\begin{figure}[t]
  \begin{center}
    \includegraphics[width=0.47\textwidth, clip, trim=5 10 40 30]{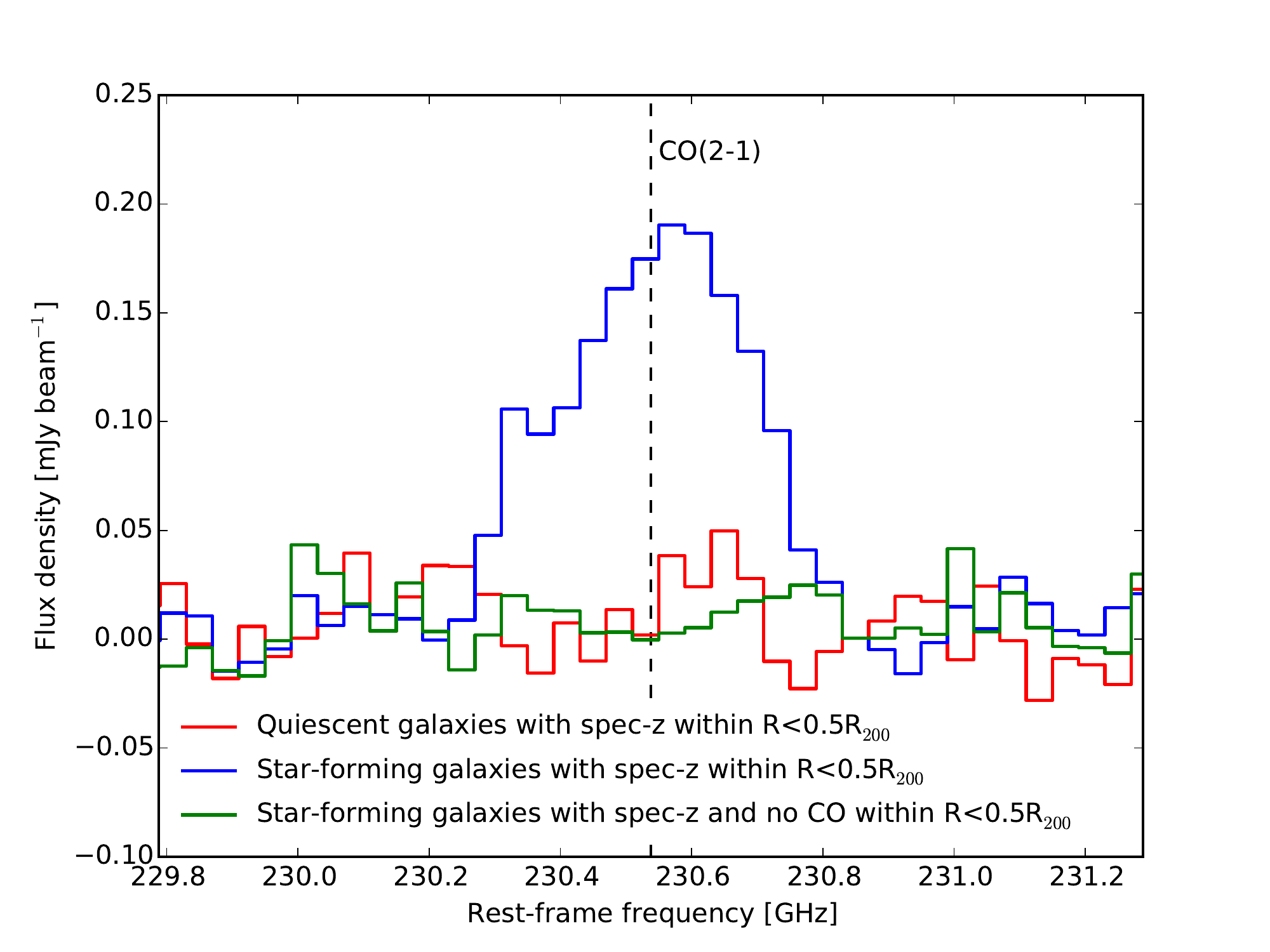}    
    \caption{
      Stacked spectra around $\nu_{\rm rest}=230.538$ GHz in ALMA Band
      3 data for quiescent galaxies (red), star-forming galaxies
      (blue), and star-forming galaxies without CO detected
      individually (green) that are spectroscopically confirmed within
      a radius of $0.5R_{200}$. The dashed line shows a frequency of
      CO(2--1) emission line. The quiescent or star-forming galaxies
      are classified by the U-V and V-J color diagram
      (Figure~\ref{fig:UVJ}). The CO(2--1) emission line is not
      detected from the stacked spectrum of quiescent galaxies. 
      The flux or the upper limit of flux of the emission line in the
      stacked spectrum is shown in Table~\ref{tbl:stack}.
      \label{fig:stack}
    }
  \end{center}
\end{figure}

The member galaxies with CO line and/or dust continuum detected are
located away from the very center of the cluster (Figure~\ref{fig:map}). 
It is worth investigating how much the gas reservoirs of the member
galaxies in the very center is left. To give a constraint on molecular
gas mass for such galaxies, the stacked data in the Band 3 are used to
discuss the average amount of the molecular gas in these member
galaxies.  

We stack the Band 3 data for two populations of quiescent galaxies and
star-forming galaxies which are classified based on the UVJ diagram
(Figure~\ref{fig:UVJ}). There are 12 quiescent galaxies and 27
star-forming galaxies that are spectroscopically confirmed within a
half of $R_{200}$. Among them, a quiescent galaxy and 15 star-forming
galaxies have CO(2--1) line detected. The redshift confirmation is
essential to shift the individual spectrum from the observed-frame to
the rest-frame. Figure \ref{fig:stack} shows the stacked spectra for
three samples from the two populations; all quiescent galaxies, all
star-forming galaxies, and star-forming galaxies without CO(2--1)
detection.  
For the sample of all star-forming galaxies, since about a half 
of them have CO(2-1) lines detected individually, the CO(2--1) line is
also detected in the stacked spectrum. On the other hand, the stacked 
spectra of the quiescent galaxies and star-forming galaxies without CO
detected individually show no detection of CO(2--1) line. Using the
intensity map integrated in velocity by 400 km s$^{-1}$, the average
CO(2--1) line flux and the 5$\sigma$ upper limit flux are estimated
for the star-forming and quiescent galaxies, respectively. We then
convert them to the molecular gas masses in the same manner as in
\S~\ref{sec:results.Mgas}. Note that we make sure the validity of our
procedure by stacking the data for the member galaxies with CO(2--1)
detected individually. The measurements from the stacked spectra are
shown in Table~\ref{tbl:stack}. 

The upper limit of molecular gas for the quiescent galaxies shows the
gas fraction of $<$ 0.08 and the depletion time scale of $<$ 9.8 Gyr,
suggesting that the quiescent galaxies in the center consume most of
their molecular gas. On the other hand, star-forming galaxies still
have enough gas to keep forming stars. 
However, many star-forming galaxies in the cluster have larger gas
reservoirs compared with the field galaxies
(\S~\ref{sec:discussion.SR}), and they also show the larger depletion 
time as they deviate from the MS. This may imply that it is difficult
to consume the molecular gas by only star formation.
The starvation to stop supply gas to the cluster galaxies as well as  
the ram-pressure to strip gas from the galaxies may be required to
effectively reduce the amount of the molecular gas, which can
accelerate the growth of cluster galaxies to become quiescent galaxies
in the cluster core. Indeed, it is observed that the molecular gas as
well as HI gas are stripped from galaxies by ram pressure in the local
galaxy cluster \citep{Sivanandam2010,Jachym2014}.  
Also, virial shock in the massive halo of this cluster at $z=1.46$
would prevent the cold gas stream from accreting to the member
galaxies settled in the cluster center
\citep{Birnboim2003,Dekel2006}. 
Other possibility is galaxy mergers. The merging of gas-rich galaxies
can induce starburst in the galaxy center \citep{Hopkins2008}, which
results in consumption of gas reservoirs. The stellar masses of the
gas-rich galaxies in this cluster are comparable to and/or a factor of
$\sim$2--5 smaller than those of quiescent galaxies. A perspective of
mass growth supports that the merging is one of possible processes.
The feedback from AGN and SN can also have a role in the quenching
mechanism. Outflows of massive molecular gas by AGNs and SNe are
observed from ultraluminous infrared galaxies (ULIRGs) in the local
Universe \citep{Feruglio2010,Sturm2011,Cicone2014}. Since there is no
evidence that a high fraction of the starburst galaxies and AGNs are
found in this cluster, the feedback may be an inadequate process for
environmental effects to the transition of star-forming galaxies to
quiescent galaxies in galaxy clusters, however it would be one of the
important processes not only to reduce the efficiency of star
formation but also to blow the gas off from the galaxies. 

There are a few previous studies to show a constraint on molecular gas
fraction for quiescent galaxies at similar redshifts. 
\citet{Sargent2015} present the upper limit of CO(2--1) luminosity in
an early-type galaxies at $z=1.43$ with IRAM/Plateau de Bure
Interferometer (PdBI), which shows the 3$\sigma$ upper limit of gas
fraction is $\la10$\%. \citet{Gobat2017} use the stacked SED ranging
from MIR to radio for about thousand early-type galaxies at 
$\langle z \rangle$=1.76 to give a constraint on their molecular gas
reservoirs. The gas mass is derived from the dust mass that is
estimated from the SED under the assumption of a metallicity-dependent
gas-to-dust ratio. They derive the gas fraction of $\sim13$\%.  
Note that the stellar masses shown in the literature is converted to
that with \citet{Chabrier2003} IMF. 
The gas fraction that we derive for the quiescent galaxies in the
galaxy cluster at $z=1.46$ is similar to those for the field quiescent
galaxies at similar redshifts, suggesting that quiescent galaxies
consume the fuel of gas down to similar low level of $\la10\%$
irrespective of the environment. However, The ATLAS$^{\rm 3D}$ project
shows that the early-type galaxies in the local Universe have a gas
fraction an order of magnitude smaller than that for the galaxies at
$z\sim1.5$ \citep{Young2011}. Preferably, the local post-starburst
galaxies seem to have a gas fraction similar to the high-$z$
cluster quiescent galaxies \citep{French2015}.

\section{Summary and conclusions}
\label{sec:conclusion}

We conduct the ALMA observations in Band 3 and Band 7 in the X-ray
galaxy cluster, XMMXCS J2215.9-1738, at $z=1.46$. While the Band 3
data allow us to detect CO(2--1) emission lines from cluster member
galaxies, the Band 7 data allow us to detect dust continuum
emissions at 870 \micron. We use these ALMA data to investigate
molecular gas reservoirs in the member galaxies within a
cluster-centric radius of $\sim$ \R200\ and then discuss the evolution
of their star formation activities in terms of star formation
efficiency and gas consumption.

\citet{Hayashi2017} already report the discovery of 17 CO(2--1)
emission lines associated with the cluster. In this paper, we newly
detect nine 870 \micron\ sources in the Band 7 data. Although one
source is a foreground galaxy, the other eight galaxies are confirmed
to be cluster member galaxies. Seven galaxies have both CO(2--1) lines
and dust continuum emissions detected, and the position of dust
continuum emission is consistent with that of CO(2--1) emission. 
Consequently, we have CO(2--1) emission lines and/or dust continuum
emissions from 18 member galaxies within $\sim R_{200}$. 
The rest-frame U-V versus V-J color diagram shows that most of the CO
lines and/or dust emissions are detected from dusty star-forming
galaxies.    

We derive molecular gas masses from the CO luminosities using the
metallicity-depended (i.e., stellar mass-depended) conversion factors
\citep{Tacconi2018} as well as dust continuum luminosities according 
to \citet{Scoville2016} while taking into account of the
metallicity-dependency of the dust-to-gas ratio. The molecular gas  
masses derived from the two ways are consistent with each other.

We investigate the gas fraction and the depletion time scale as a
function of stellar mass, offset from the main sequence of
star-forming galaxies, cluster-centric radius, and accretion
phase. The galaxies with larger SFRs at a given stellar mass show a
larger gas fraction and a smaller depletion time scale. There is no
strong dependence of gas fraction and depletion time on the
cluster-centric radius and the accretion phase. The cluster member
galaxies with CO and/or dust continuum detected tend to have a larger
gas fraction and a larger depletion time, compared with those from the
scaling relations for field galaxies. If infalling regions and
filaments around galaxy clusters help feed the gas through inflow to
member galaxies, the cluster galaxies can have the larger gas
reservoirs than the field galaxies at $z\sim1.5$. Nevertheless, the
cluster galaxies must become more inefficient in star formation than
field galaxies. As the member galaxies are infalling to closer to the
center, the deviation of gas fraction and depletion time from the
scaling relations seems to get larger. Therefore, some environmental
effects peculiar to galaxies associated with galaxy clusters may
reduce the efficiency of star formation. 

Massive quiescent galaxies in the cluster core no longer have large
gas reservoirs and efficient star formation. We stack the Band 3
spectra for 12 quiescent galaxies within a radius of 0.5 \R200. 
However, no CO(2--1) emission line is detected from the stacked
spectrum. The upper limits of molecular gas and molecular gas fraction
are estimated to be $\la10^{10}$ \Msun\ and $\la10$\%, respectively,
which are similar to those for quiescent galaxies in general fields at
similar redshifts. This suggests that irrespective of the environment,
the massive quiescent galaxies consume most of the fuel of gas and
evolve passively in the center of the cluster. 
We speculate that since cluster member galaxies are subject to
additional environmental effects such as ram-pressure, starvation, and
merging compared with field galaxies, cluster galaxies is easier to
reduce gas reservoirs and then quench star formation, which results in
a larger fraction of quiescent galaxies in galaxy clusters rather than
in general fields.



\acknowledgments
We thank the anonymous referee for providing constructive comments and
suggestions.
MH acknowledges the financial support by JSPS Grant-in-Aid for Young
Scientists (A) Grant Number JP26707006 and was also supported by the
ALMA Japan Research Grant of NAOJ Chile Observatory, NAOJ-ALMA-180.
KK acknowledges the financial support by the JSPS Grant-in-Aid for
Scientific Research (A) JP25247019.
YY is thankful for the JSPS fellowship.
This paper makes use of the following ALMA data:
ADS/JAO.ALMA\#2011.1.00623.S and ADS/JAO.ALMA\#2015.1.00779.S. 
ALMA is a partnership of ESO, NSF (USA) and NINS (Japan), together
with NRC (Canada), NSC and ASIAA (Taiwan), and KASI (Republic of
Korea), in cooperation with the Republic of Chile. The Joint ALMA
Observatory is operated by ESO, AUI/NRAO and NAOJ.
This work uses the data collected at Subaru Telescope, which is
operated by the National Astronomical Observatory of Japan. 
This work also uses the data based on observations obtained with
MegaPrime/MegaCam, a joint project of CFHT and CEA/IRFU, at the
Canada-France-Hawaii Telescope (CFHT) which is operated by the
National Research Council (NRC) of Canada, the Institut National des
Science de l'Univers of the Centre National de la Recherche
Scientifique (CNRS) of France, and the University of Hawaii. This work
is based in part on data products produced at Terapix available at the
Canadian Astronomy Data Centre as part of the Canada-France-Hawaii
Telescope Legacy Survey, a collaborative project of NRC and CNRS.  
Some of the data presented in this paper were obtained from the
Mikulski Archive for Space Telescopes (MAST). STScI is operated by the
Association of Universities for Research in Astronomy, Inc., under
NASA contract NAS5-26555. Support for MAST for non-HST data is
provided by the NASA Office of Space Science via grant NNX09AF08G and
by other grants and contracts

{\it Facilities:} \facility{ALMA, Subaru, CFHT, HST}.


\end{document}